# Stability and magnetism in complex multivacancy graphene systems

Ricardo Faccio[1,2] and Alvaro W. Mombrú[1,2,3]

1 Centro NanoMat, Cryssmat-Lab, DETEMA, Polo Tecnológico de Pando, Facultad de Química, Universidad de la República, Cno. Saravia s/n, CP 91000, Pando, Uruguay

2 Centro Interdisciplinario en Nanotecnología, Química y Física de Materiales, Espacio Interdisciplinario, Universidad de la República, Montevideo, Uruguay.

3 Author to whom all correspondence should be addressed

E-mail: amombru@fq.edu.uy

**Abstract**

This work is focused in the stability and magnetic properties of complex graphene multivacancy systems, studied by Density Functional Theory calculations. The removal of a sequence of carbon atoms -i.e. the complementary figure- leads to these multivacancy systems. We show evidence that dendritic-like complementary figures yield to very stable defected graphene structures. If this dendritic complementary figure is aliphatic, the stress of the defected graphene system is enough to decrease the symmetry of such a system. In the cases that the complementary figure is based in phenyl- or pyrenyl-like cores, the remaining system with the vacancy will preserve the symmetry of the removed segment yielding to flower-like structures after optimization. This has consequences in the magnetic arrangement based on the dangling bonds located at the loose vertices of the pentagonal rings. The cases with orthogonal anisotropy, the magnetic order is driven by the symmetry of the vacancy system -i.e. the symmetry of the complementary figure-.

**Introduction**

The importance of carbon nanomaterials has been established in the last decades due to the impact on materials development: fullerenes [1], carbon nanotubes [2], graphite [3-7] and graphene [8], research on edge effects in graphene, such as nanoribbons [9] and vacancies [10-16].

In particular, research is being carried out on graphene multivacancy systems, with focus in the occurrence of magnetism due to the presence of dangling bonds [17] and the stability related to the configuration of the vacancies [18]. These studies introduce the concept of the complementary figure, i.e. the geometric figure of the atomic arrangement that is extracted from graphene when the multivacancy is created, and the usefulness of it to perform more accessible interpretations and to deduce projections for different vacancies configurations.

In fact vacancies are features that should be studied in depth due to the modifications in the physical properties that their presence induces as already seen for carbon nanotubes [19] and graphene itself [17].

Although previous work has been performed about stability on multivacancy systems [18], it was focused on zigzag and armchair and other simple aliphatic-like complementary figures.

This manuscript is intended to go further with this subject by calculating the stability on more complex multivacancies systems, studying them through the first principles-based calculations of the formation energies of these defected systems. In order to achieve this task, we performed Density Functional Theory (DFT) calculations [20,21].

This subject is becoming very relevant due to the recent advances in vacancies creation techniques [22-24].

It should be mentioned that the scope of this manuscript is disregarding the approach of reconstructed vacancies, 585 and 555-777 double vacancies or Stone Wales defects [25,26], just focusing in the vacancies formation without any rearrangement.

## 2. Computational Methods

Periodic spin polarized band structure calculations were performed on graphene multivacancies systems with the use of the DFT program VASP (Vienna *ab initio* simulation package) [27]; pseudopotentials were applied with a plane-wave basis. The exchange correlation potential was chosen as the generalized gradient approximation (GGA) [28] in a projector augmented wave (PAW) method [29,30]. An energy cutoff of 500 eV was used to expand the Kohn–Sham orbitals into plane wave basis sets. In all of the calculations reported here the k-point mesh was taken equivalent to 4x4x1 for the full (reducible) Brillouin Zone, allowing the convergence of total energy, stress components and ionic forces. Both supercell dimensions and ions positions were allowed to optimize, until residual forces and stress tensor components were positioned down to 0.01 eV/Å and 5 kbar respectively.

## 3. Discussion and Results

As in a previous work [17], the formation energies per carbon atom, showing the stability in these multivacancy systems, were calculated in the following way:

$E(k) = E_k - NE_1$

where N is the number of carbon atoms in the graphene supercell after the vacancy has been created -i.e. the complementary figure has been removed-, k is the number of carbon atoms extracted to create the vacancy, $E_k$ is the energy of the optimized structure for a k-order vacancy and $E_1$ is the energy per carbon atom in regular graphene. The nomenclature *z*k and *a*k stands for zigzag and armchair construction of the complementary figure and *o*k for other configurations (k is the order of the vacancy). In this work only *o*k vacancies are studied. Additional number n as in *o*kn is just an ordinal number to identify different configurations among k-order vacancy systems.

Figure 1 shows the cost of removing an extra atom from a branched 5-order vacancy system to yield 6-order vacancies with different configurations, depending on the place where the atom is extracted (the representation is performed through the complementary figure already defined and previously used [17,18]). Previous work has explained the different values for all the cases, but the o66 case is not shown. The negative value of the subtraction E(*o*66)-E(*o*51) deserves a special consideration.

When going from *o*51 to *o*66, there is no longer any dangling bond and instead, four pentagons are formed. This can be seen in figure 2 b). According to the rules previously defined [14], terminal carbon atoms at the complementary figure, imply the formation of pentagons in the defected graphene structure that remains after the removal. In the case of o66 there are no vertices exposed in the complementary figure, so no dangling bond should occur after the vacancy creation, according to the same rules.

We found this extra stabilization after the disappearance of dangling bonds and the armonic occurrence of pentagons in the defected graphene structure. The absence of exposed vertices and the proliferation of terminal carbon atoms in the complementary figure, is very characteristic of a dendritic-like figure.

In order to evaluate our hypothesis about the cause for the stability of the *o*66 configuration and the dendritic character that could act as a stabilizing factor for the formation of multivacancies, we tried to study other vacancies with dendritic-like complementary figures.

It is clear that the symmetric 4-order vacancy, shown in figure 2 a), already studied from the magnetic point of view in [17], matches the characteristics of a dendritic-like complementary figure. As a matter of fact, it is the smallest possible multivacancy with this character.

The energy of this system is 9,566 eV, lower than the other 4-order vacancy systems, the zigzag (10,793 eV) and armchair (14,103 eV) ones, what would suit with our hypothesis. This effect can be seen in figure 3.

After this, we tried to check the validity of the hypothesis of the dendritic-like complementary figure, as a general trend for stable multivacancy systems.

For this reason we studied multivacancies with dendritic-like complementary figures (i.e. with no exposed vertices, and many terminal carbon atoms), for 8-, 18-, and 30- order vacancies.

They are shown in figures 4, 5 and 6, respectively: the complementary figure and the defected graphene structure after it has been removed, explicitly exhibited, and the multivacancy system after optimization.

The 8-order vacancy system is obtained when an aliphatic-like complementary figure is extracted. Although this complementary figure is symmetric, as can be seen in figure 4 a), after optimization, the symmetry is broken. This effect is due to the close position of the terminal carbon atoms, labeled as *a* and *b*. This two terminal atoms removed, generate a competition for the pentagon occurrence, either where *a* or *b* was located, yielding to a clearly non-symmetric vacancy after the optimization of the structure. This kind of competition when a multivacancy system is formed is new to our knowledge and leads to unexpected results, that were not included in the rules previously postulated in [17]. This 8-order vacancy shows a net magnetic moment, in spite of the absence of exposed vertices in the complementary figure, as can be seen in Table 1, where formation energies per carbon atom and magnetic moments for the systems here presented are exhibited. The non-symmetry of the vacancy allows the occurrence of a magnetic system where the spin-up and spin-down charge density in the hole is driven only by the alternation in consecutive atoms, as can be seen in figure 4 b). The difference between spin-up and spin-down population is shown in figure 4 c).

In order to achieve higher order vacancies that follow the dendritic character, complementary figures based in a phenyl-like structure, were designed. In the case of the 18-order vacancy, the complementary figure has $D_{3h}$ symmetry, and the competition of terminal carbon atoms is present too. As can be seen in figure 5 a), the $x_2$ terminal atoms are in clear competition against $x_1$ and $x_3$, for pentagon formations when optimization is processed (x = a, b and c). The result after the optimization can be seen in figure 5 b), and even though in the complementary figure there is no exposed vertex -i.e. the idea of a dentritic structure-, the final carbon defected graphene system still shows a net magnetic moment as can be seen in Table 1. The occurrence of dangling bonds on the exposed vertices of the pentagonal rings formed after the optimization through a novel pentagonal configuration, resembling the pyrrole rings configuration in porphyrins, not previously seen in simpler, non-dendritic complementary figure based multivacancy systems (see figure 5 c)).

In this case, the phenyl-like core relieves the stress that the removal of the complementary figure causes, yielding a vacancy system that holds the symmetry of this figure. The ferromagnetic structure that arises due to the spin polarization of the charge density is driven by the alternation in the hole of the system as well as the preserved symmetry of the system, due to the already mentioned low stress.

In an effort to study higher order vacancy systems, we designed a 30-order one, which is shown in figure 6 a). As can be seen, the complementary figure is based on a pyrenyl-like core, with aliphatic terminal segments to yield the dendritic character. This complementary figure correspond to the $D_{2h}$ symmetry, which is centrosymmetric if constrained to the plane, introducing an axial anisotropy. In this case, the $x_1$ and $x_2$ terminal carbons of the complementary figure, are in competition for the formation of pentagonal rings, due to their close vicinity (x = a, b). In the orthogonal direction, the central $y_2$ terminal carbon atoms are in competition with the $y_1$ and $y_3$ atoms (y = c, d). The pyrenyl-like core of the complementary figure relieves the stress in the vacancy optimization, and the final multivacancy system keeps the symmetry of the complementary figure (figure 6 b)). The magnetic moment for this system is 4,89 $\mu_B$, corresponding to a spin-up density based on the dangling bonds at each loose vertex of the pyrrole-like pentagonal rings. If this magnetic arrangement was only driven by the alternation of spin character in successive carbon atoms, the final structure should exhibit no net magnetic moment, and four dangling bonds with spin-up arrangements at the left of figure 6 c) would oppose the other four dangling bonds with spin-down configuration at the right. This possibility is ruled out by symmetry considerations, since the vacancy has the same symmetry as the complementary figure due to the eccentricity of the hole formed by the strong influence of the pyrenyl-like core. The centrosymmetric point group that result from removing such complementary figure leads to the need of a full spin-up configuration to keep the invariance of the magnetic moment through the symmetry operations that originates that group [31,32]. The anisotropy in orthogonal directions of this example, offers new insights of the magnetic behavior of the multivacancy systems, and how the symmetry of the complementary figure rules on the magnetic arrangements in the dangling bonds system against the alternation of spin in successive carbon atoms, when a minimal eccentricity of the remaining hole is achieved -i.e. there is a stress relief caused by the aromatic core in the dendritic-like complementary figure.

In order to check this new concept, we designed a 16-order vacancy system, as shown in figure 7 a). The complementary figure shows a $D_{2h}$ symmetry, with axial anisotropy, as in the already discussed 30-order vacancy system. For this reason, this example could be a good test to the key role of symmetry when orthogonal anisotropy is present in the complementary figure. In this case the competition for pentagonal ring formation is among $x_1$ and $x_2$ positions of the terminal carbon atoms of the complementary figure (x = a, b, c, d).

The phenyl-like core, as in other cases, is providing relief against the stress of the atoms removal on the vacancy formation. As a consequence, the final optimized defected carbon graphene keeps the $D_{2h}$ symmetry, which is centrosymmetric (see figure 7 b). There is a net magnetic moment after the optimization, as can be seen in Table 1, which is not expected in terms of the alternation of the spin character in adjacent carbon atoms in the hole. If the magnetic arrangement was driven by this effect, two subsets of opposite spins would occur and the calculated magnetic moment clearly imply that it is not the case. As in the 30- order vacancy case, the symmetry of the defected graphene, with a center of inversion that drives the invariance of the magnetic moments on each dangling bond to present only spin-up character, yields to a ferromagnetic local arrangement at the edge of the vacancy (see figure 7 c).

Figure 8 presents the partial density of states for the $o$66, $o$8, $o$16, $o$18 and $o$30 cases. As expected and already mentioned, the only case where there is no spin polarization corresponds to $o$66, where the density of states looks symmetric along the energy axis, with σ- and π-states contributions at the Fermi level. In the case of $o$8, $o$16, $o$18 and $o$30, the densities of states look asymmetric as consequence of the spin polarization, where σ- and π-states contributions are very different. In one hand, σ-states present more localization due to presence of atom vacancies, establishing a pseudo gap of about 1.8 eV. On the other hand, the π-states contribute in a wider energy range in comparison with σ-states, this is a consequence of the better $p_z$ hybridization between carbon atoms close to the edge of the multivacancy; which additionally present an important contribution of states at the Fermi level.

**Conclusions**

This manuscript shows some cases of complex multivacancy systems, many of them optimized to flower-like defected graphene structures, very stable and with fullerene resemblance in the structural role that pentagonal rings and their conjugation with hexagonal ones exhibit.

The dendritic-like complementary figures yield to the most stable multivacancy systems configurations studied so far. This would be in agreement with recent results [23] that show that the dendritic growth of vacancies in graphene by etching, in the macroscale, that could be driven by this preferred route from the nanoscale. The aliphatic dendritic complementary figures could lead to distortions that could decrease the symmetry of the optimized final vacancy structure. The removal of a dendritic complementary figure based on a phenyl- or pyrenyl-like core, suggested to contribute with a eccentricity of the vacancy that could relieve the stress of the defected structure. The consequence of such complementary figure is the preservation of the symmetry of this figure after the optimization process. We found that it plays a key role in the arrangement of the spins located at the dangling bonds at the vacancy. We designed two orthogonal anisotropic cases to check the importance of this factor on the final magnetic moment of the systems.

Due to the stress that the competition for the formation of pentagonal rings in adjacent positions, multivacancy systems constructed from the removal of dendritic-like complementary figures, are the exception to the rules previously established [17].

**Acknowledgments**

The authors wish to thank the Uruguayan funding institutions, CSIC, ANII and PEDECIBA.

| Vacancy system | 3 | *o*66 | *o*8 | *o*16 | *o*18 | *o*30 |
|---|---|---|---|---|---|---|
| E(k) (eV) | 8,194 [18] | 13,153 | 18,138 | 28,101 | 29,341 | 39,802 |
| Magnetic Moment ($\mu_B$) | 1,04 [17] | 0 | 1,16 | 2,34 | 6,21 | 4,89 |

Table 1. Formation energy per carbon atom and magnetic moment for defected graphene structures.

**FIGURE CAPTIONS**

Figure 1. Effect of the position where a vacancy is grown from a 5- to a 6- order one. The figures below are the E(*o*kn) energies. The defected graphene is omitted and the spin density configurations are not shown for clarity. The analysis of the non-dendritic cases -*o*61, *o*62, *o*63, *o*64-, as well as *o*65, is presented in [18]. The *o*66 case is framed and the energy difference (E*o*66-E*o*51) is shown with an arrow. Energies in eV.

Figure 2. a) 4- and b) 6- order vacancy systems (up) with complementary figures dendritic shapes (down).

Figure 3. Formation energy per carbon atom as a function of the order of the vacancy. Square and circle symbols correspond to zigzag and armchair complementary figures, previously discussed in [18] and diamonds symbols, to dendritic complementary figures. The thick arrow shows the trend of the armchair complementary figure, to make clear how the dendritic structures are more stable for increasing order. The solid line shows the linear fit for the dendritic structures ($R^2 = 0.96$).

Figure 4. a) Complementary figure for 8- order vacancy, b) vacancy after optimization and c) differences between the spin-up and spin-down populations.

Figure 5. a) Complementary figure for 18- order vacancy and b) vacancy after optimization and c) differences between the spin-up and spin-down populations.

Figure 6. a) Complementary figure for 30- order vacancy and b) vacancy after optimization and c) differences between the spin-up and spin-down populations.

Figure 7. a) Complementary figure for 16- order vacancy and b) vacancy after optimization, and c) differences between the spin-up and spin-down populations.

Figure 8. Partial density of states for: *o*66, *o*8, *o*16, *o*18, *o*30; indicating σ and π contributions. Blue and red lines represents π- and σ-states respectively.

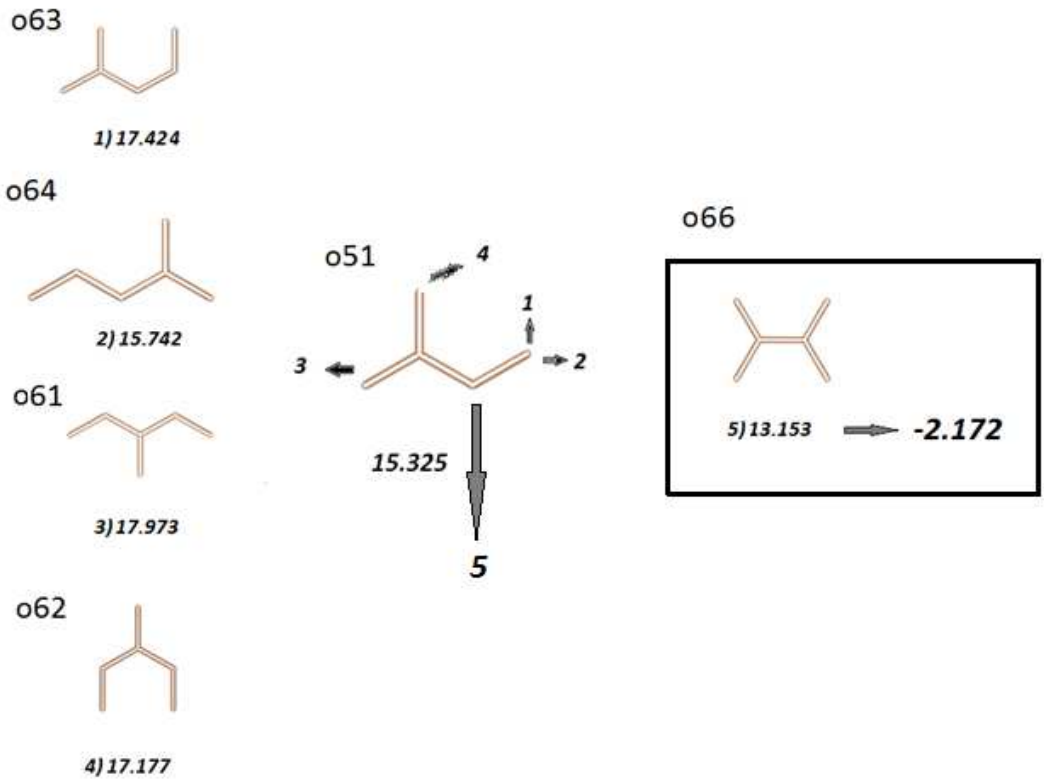

Figure 1. Effect of the position where a vacancy is grown from a 5- to a 6- order one. The figures below are the E(*o*kn) energies. The defected graphene is omitted and the spin density configurations are not shown for clarity. The analysis of the non-dendritic cases -*o*61, *o*62, *o*63, *o*64-, as well as *o*65, is presented in [18]. The *o*66 case is framed and the energy difference (E*o*66-E*o*51) is shown with an arrow. Energies in eV.

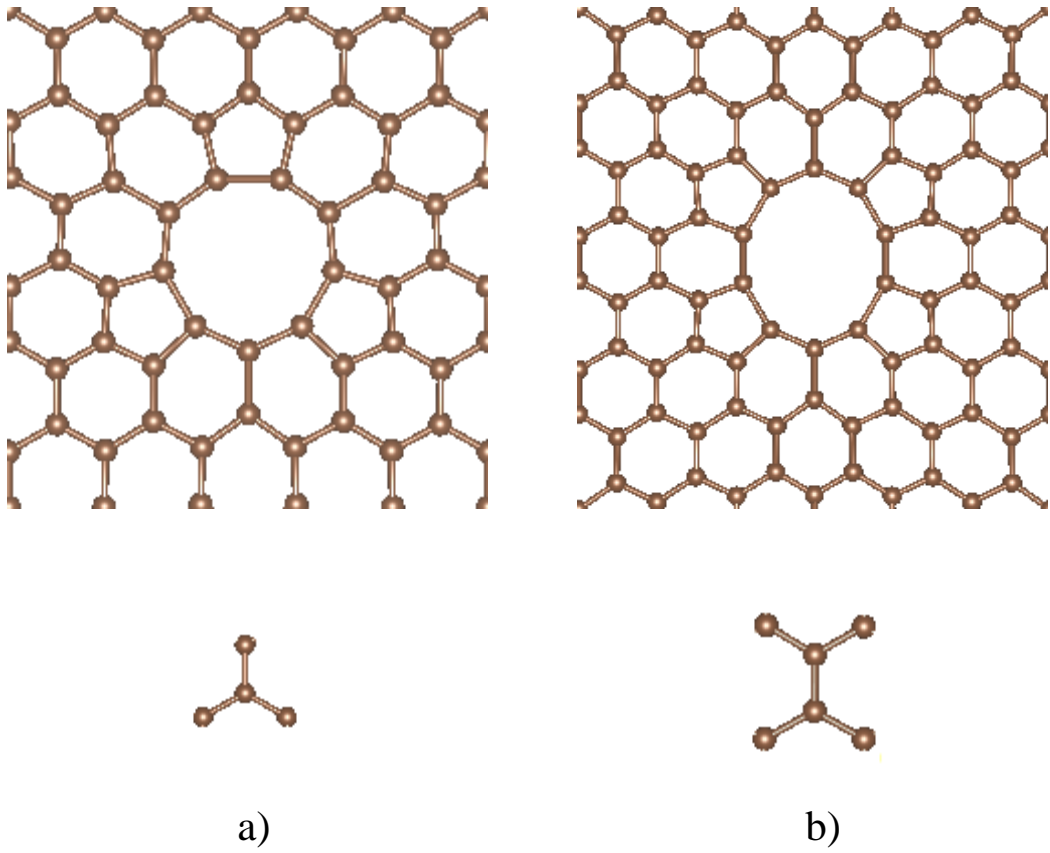

Figure 2. a) 4- and b) 6- order vacancy systems (up) with complementary figures dendritic shapes (down).

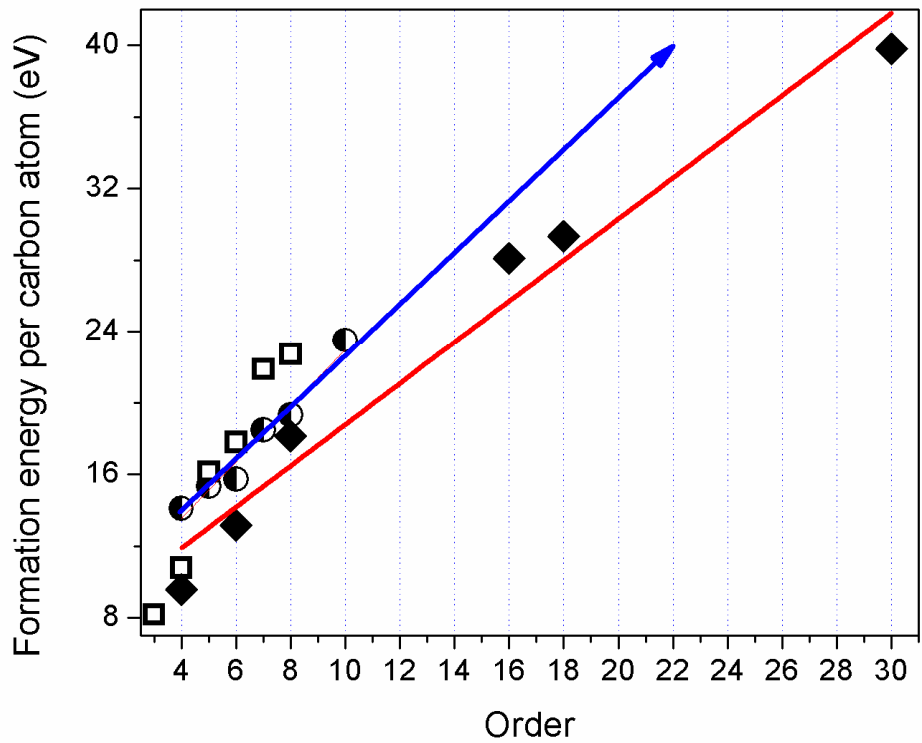

Figure 3. Formation energy per carbon atom as a function of the order of the vacancy. Square and circle symbols correspond to zigzag and armchair complementary figures, previously discussed in [18] and diamonds symbols, to dendritic complementary figures. The thick arrow shows the trend of the armchair complementary figure, to make clear how the dendritic structures are more stable for increasing order. The solid line shows the linear fit for the dendritic structures ($R^2 = 0.96$).

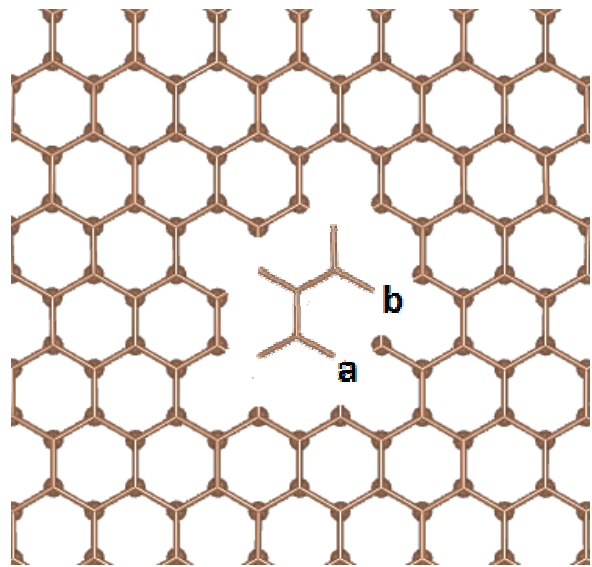

a)

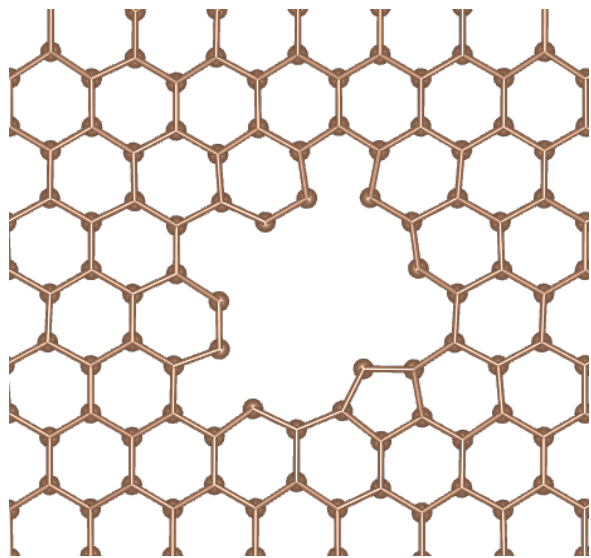

b)

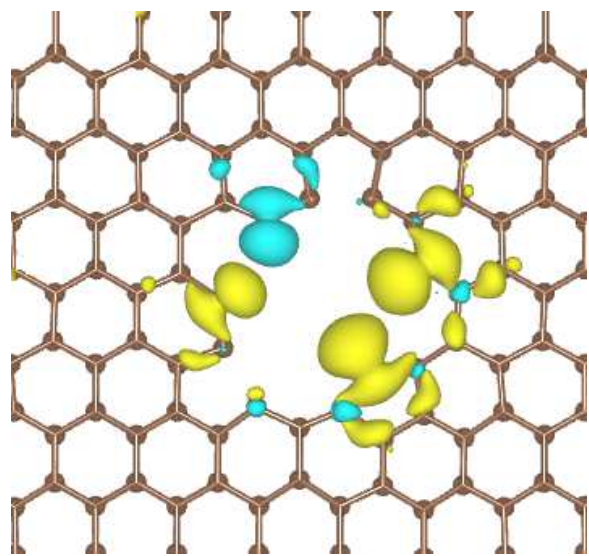

c)

Figure 4. a) Complementary figure for 8- order vacancy, b) vacancy after optimization and c) differences between the spin-up and spin-down populations.

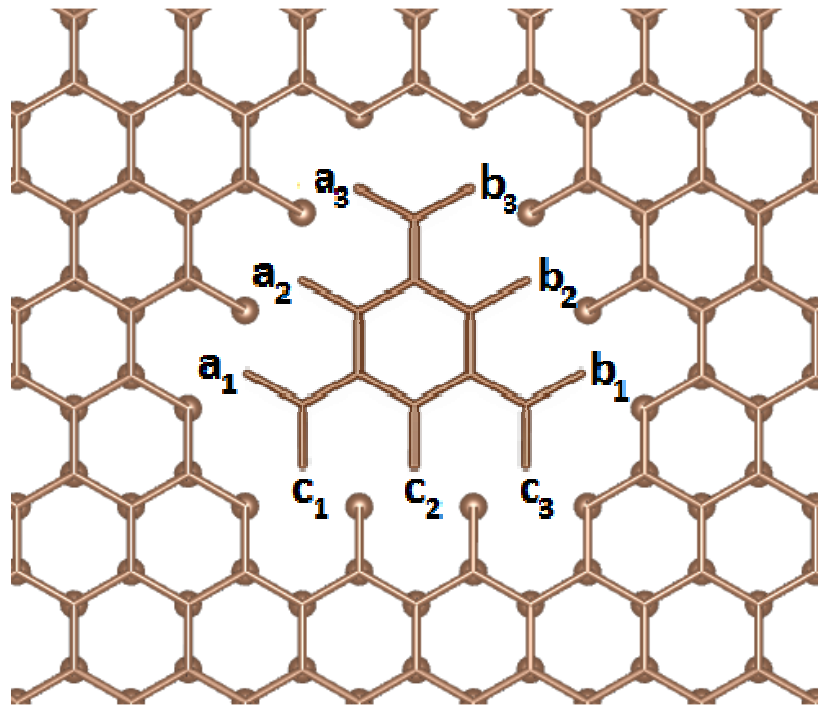

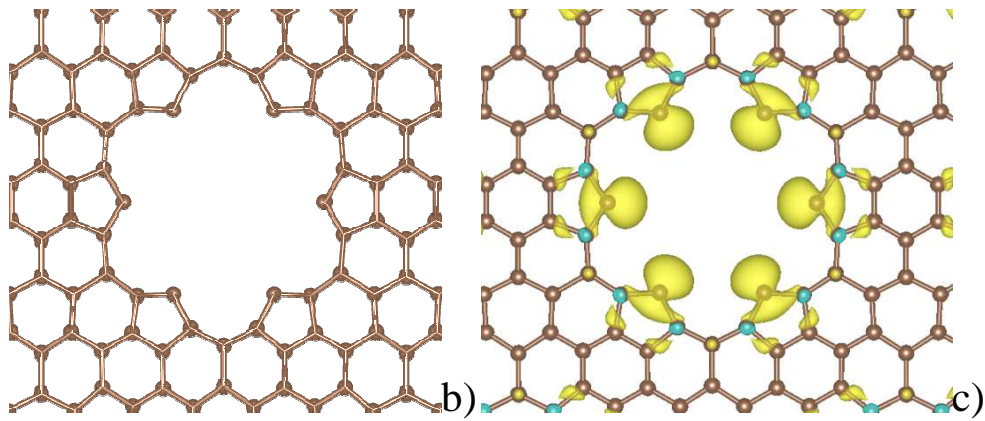

Figure 5. a) Complementary figure for 18- order vacancy and b) vacancy after optimization and c) differences between the spin-up and spin-down populations.

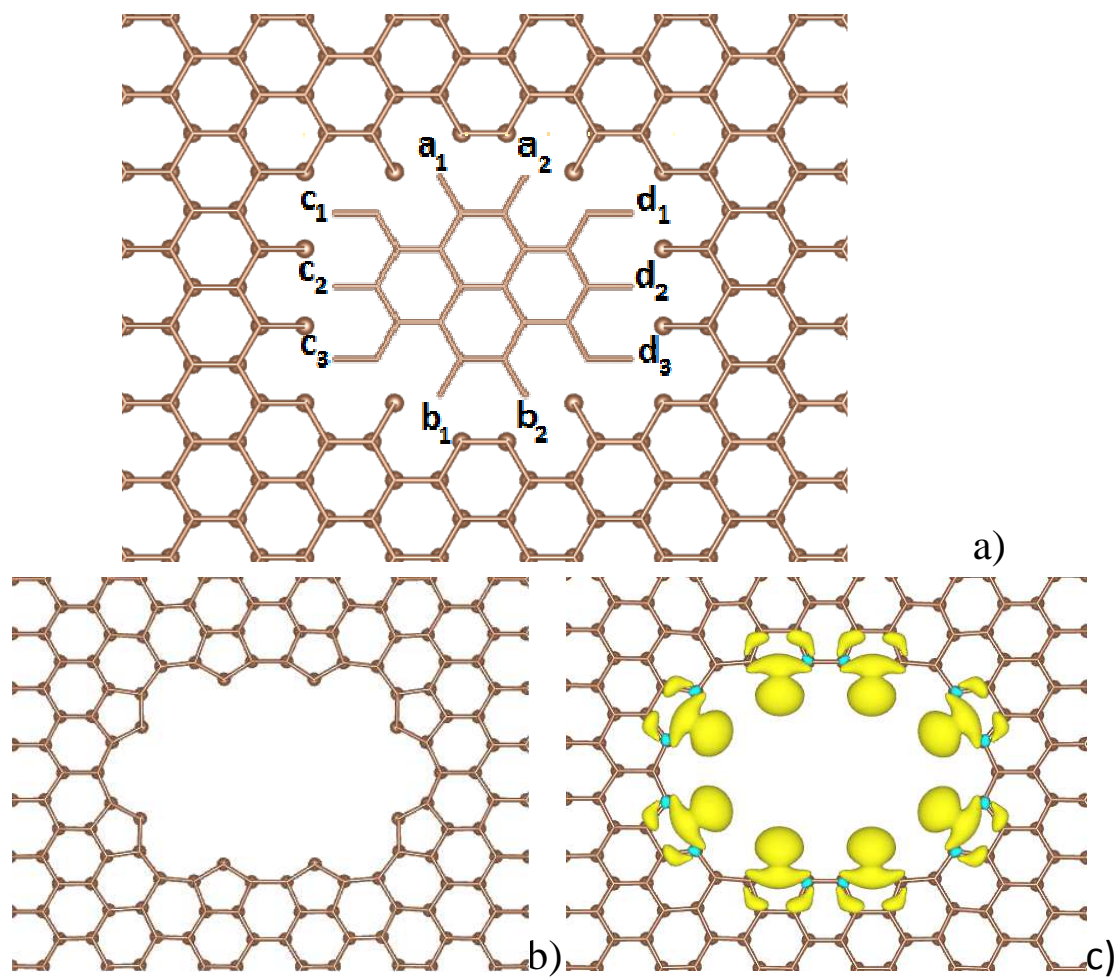

Figure 6. a) Complementary figure for 30- order vacancy and b) vacancy after optimization and c) differences between the spin-up and spin-down populations.

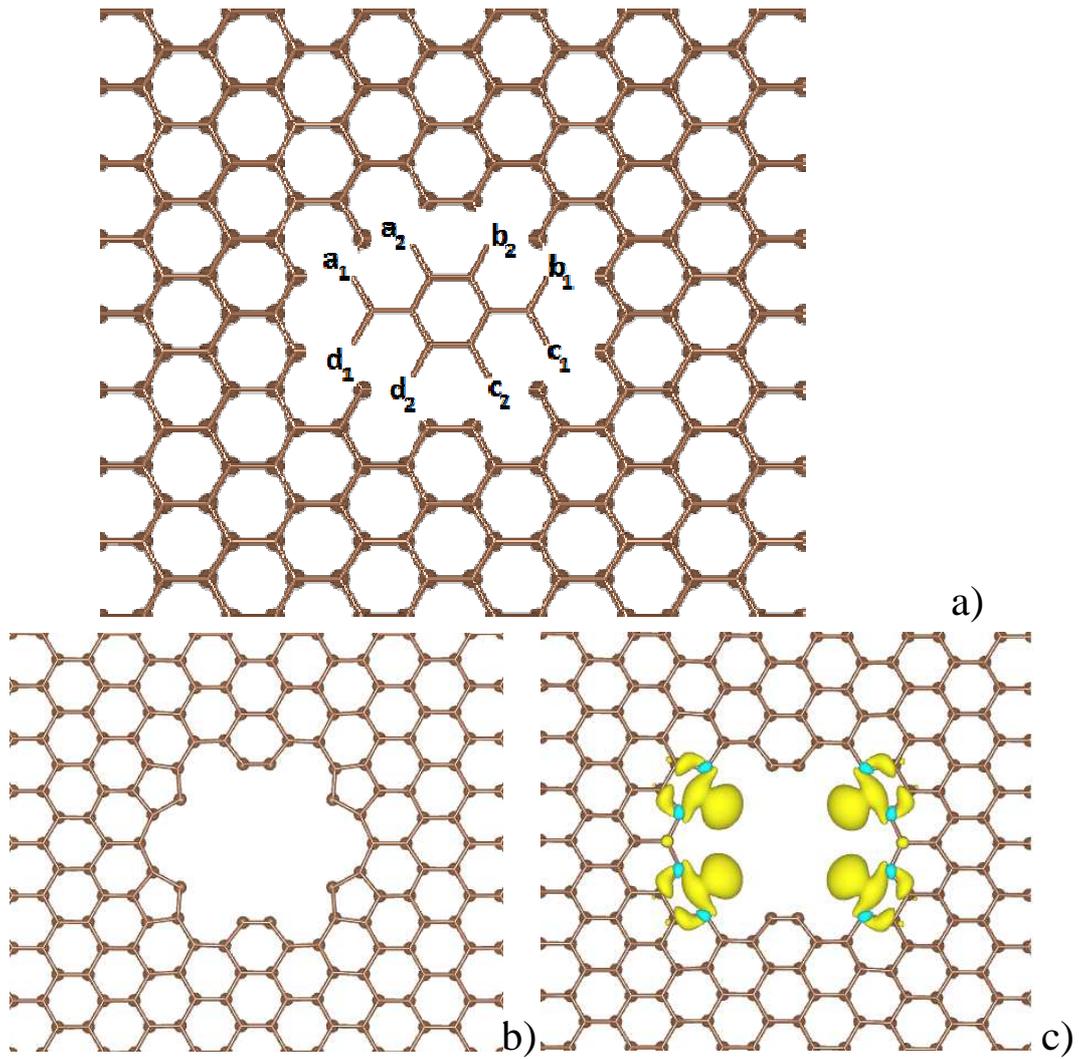

Figure 7. a) Complementary figure for 16- order vacancy and b) vacancy after optimization, and c) differences between the spin-up and spin-down populations.

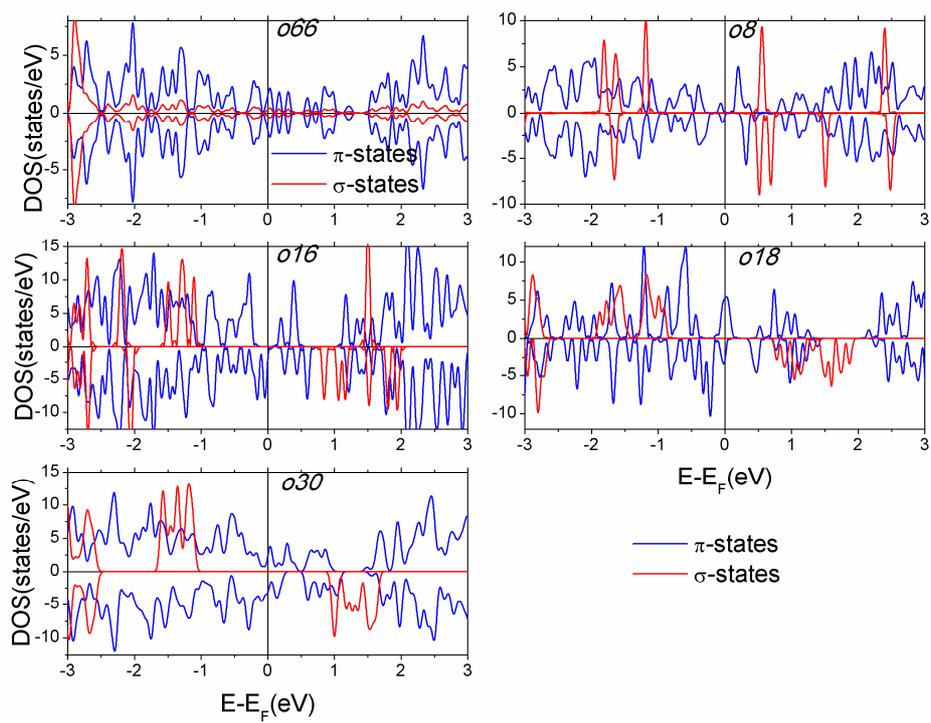

Figure 8. Partial density of states for: *o*66, *o*8, *o*16, *o*18, *o*30; indicating σ and π contributions. Blue and red lines represents π- and σ-states respectively.